\documentclass[onecolumn,showpacs,preprintnumbers,amsmath,amssymb]{revtex4}
 \usepackage{graphicx}
\usepackage{latexsym}
\usepackage{dcolumn}
\usepackage{stmaryrd}

\usepackage{graphicx}
\usepackage{dcolumn}
\usepackage{bm}

\usepackage{bm}

\begin{document}
\newcommand{\eq}{\begin{equation}}                                                                         
\newcommand{\eqe}{\end{equation}}             

\title{Heat conduction: a telegraph-type model with \\ self-similar 
behavior of solutions II} 

\author{ I. F. Barna$^a$ and R. Kersner$^b$}
\address{$^a$ KFKI Atomic Energy Research Institute of the Hungarian Academy 
of Sciences, \\ (KFKI-AEKI), H-1525 Budapest, P.O. Box 49, Hungary, \\ 
 $^b$University of P\'ecs, PMMK, Department of Mathematics and Informatics, 
Boszork\'any u. 2, P\'ecs, Hungary}

\date{\today}

\begin{abstract} 
In our former study (J. Phys. A: Math. Theor. 43, (2010) 325210  or arXiv:1002.0999v1 [math-ph]) 
we introduced a modified Fourier-Cattaneo law and derived a non-autonomous 
telegraph-type heat conduction equation which has desirable self-similar 
solution. Now we present a detailed in-depth analysis of this model  
and discuss additional analytic solutions for different parameters. The solutions 
have a very rich and interesting mathematical structure due to various 
special functions.   
\end{abstract}

\maketitle
The heat equation propagates perturbation with infinite velocity, which is a well-known
theoretical problem from a long time. A historical review of the different way-outs can be found in \cite{gurt,jos1,jos2},
According to Gurtin and Pipkin \cite{gurt,jos1,jos2}, the most general form of the flux in  linear heat conduction and diffusion related to the flux $q$  expressed in one space dimension via an integral over the history of the temperature gradient
\begin{equation}\label{1}
  q = -\int_{-\infty}^{t}Q(t-t')\frac{\partial T(x,t')}{\partial x}dt'
\end{equation}
 where $Q(t-t')$ is a positive, decreasing relaxation function that tends to zero
as $t-t' \rightarrow  \infty $ and $T(x,t)$ is the temperature distribution. 

There are two notable relaxation kernel functions: if $Q_1(s) = k\delta(s)$ where $\delta(s)$ is a Dirac delta "function", then we will get back the original  Fourier law.

In out former study \cite{imi} we introduced the $Q(t-t')=\frac{k\tau^l}{(t-t'+\omega)^l}$
time dependent kernel which leads to a non-autonomous telegraph-equation 
\begin{equation}\label{11}
\epsilon\frac{\partial^2 T(x,t)}{\partial t^2} +
\frac{a}{t} \frac{\partial T(x,t)}{\partial t} = \frac{\partial^2
T(x,t)}{\partial x^2} 
\end{equation}
which has self-similar solution of the form of 
\begin{equation}
T(x,t)=t^{-\alpha}f\left(\frac{x}{t^\beta}\right):=t^{-\alpha}f(\eta).
\end{equation}. 
  
The similarity exponents $\alpha$ and $\beta$ are of primary physical importance since 
$\alpha$  represents the rate of decay of the magnitude $T(x,t)$, 
while $\beta$  is the rate of spread 
(or contraction if  $\beta<0$ ) of the space distribution as time goes on.  
Substituting this into (\ref{11}): 
\begin{eqnarray}\label{12}
f''(\eta) t^{-\alpha -2} [\epsilon\beta^2 \eta^2] + \nonumber \\
f'(\eta) \eta t^{-\alpha -2} [\epsilon\alpha\beta  -
\epsilon\beta(-\alpha-\beta-1)  - \beta a] + \nonumber \\
f(\eta) t^{-\alpha -2} [-\epsilon\alpha(-\alpha-1) - a\alpha] =
f''(\eta) t^{-\alpha -2\beta}
\label{3},
\end{eqnarray}
where prime denotes differentiation with respect to $\eta.$

One can see that this is an ordinary differential equation(ODE)
if and only if $\alpha + 2 = \alpha + 2\beta$ ({\it {the universality relation}}).
So it has to be 
\begin{equation}
\beta =1  
\end{equation} 
while $\alpha$ can be any number. The corresponding ODE is: 
\begin{equation}\label{13}
    f''(\eta) [\epsilon\eta^2-1 ] +
f'(\eta) \eta(2\epsilon\alpha  + 2\epsilon -a) +
f(\eta) \alpha (\epsilon \alpha+\epsilon- a) = 0.
\end{equation}

In pure heat conduction-diffusion processes - no sources  or sinks - the heat mass is
conserved: the integral of $T(x,t)$ with respect to $x$ does not depend on time $t$.
For $T(x,t)$ this means
\begin{equation}
\int T(x,t) dx=t^{-\alpha} \int f(\frac{x}{t})dx=t^{-\alpha +1}\int f(\eta)d\eta=const
\end{equation}
if and only if $\alpha=1.$
Eq. 6 can be written as
\begin{equation}
(\varepsilon f\eta^2-f)''=a(\eta f)'
\end{equation}
which after integration and supposing $f(\eta_0)=0$ the first integration constant 
is zero for some $\eta_0$ gives
\begin{equation}\label{14}
    \frac{df}{f}=\frac{a\eta d\eta}{\varepsilon \eta^2-1}.
\end{equation}
The solution which is globally bounded and positive
in the domain $\{(x,t): 1-\varepsilon \eta^2 >0 \}$ and has the form \begin{equation}
f=(1-\varepsilon \eta^2)^{\frac{a}{2\varepsilon}-1}_+
\end{equation} 
(With substitution and direct derivation it can be seen that $f=(\varepsilon \eta^2-1)^{\frac{a}{2\varepsilon}-1} $ is a solution also.) 
The corresponding self-similar solution is
\begin{equation}\label{15}
    T(x,t)=\frac{1}{t}\left(1-\varepsilon \frac{x^2}{t^2}\right)^{\frac{a}{2\varepsilon}-1}_+
\end{equation}
which was presented in our former study \cite{imi} with a detailed analysis. 
 
\begin{figure}*
\scalebox{0.65}{
\rotatebox{0}{\includegraphics{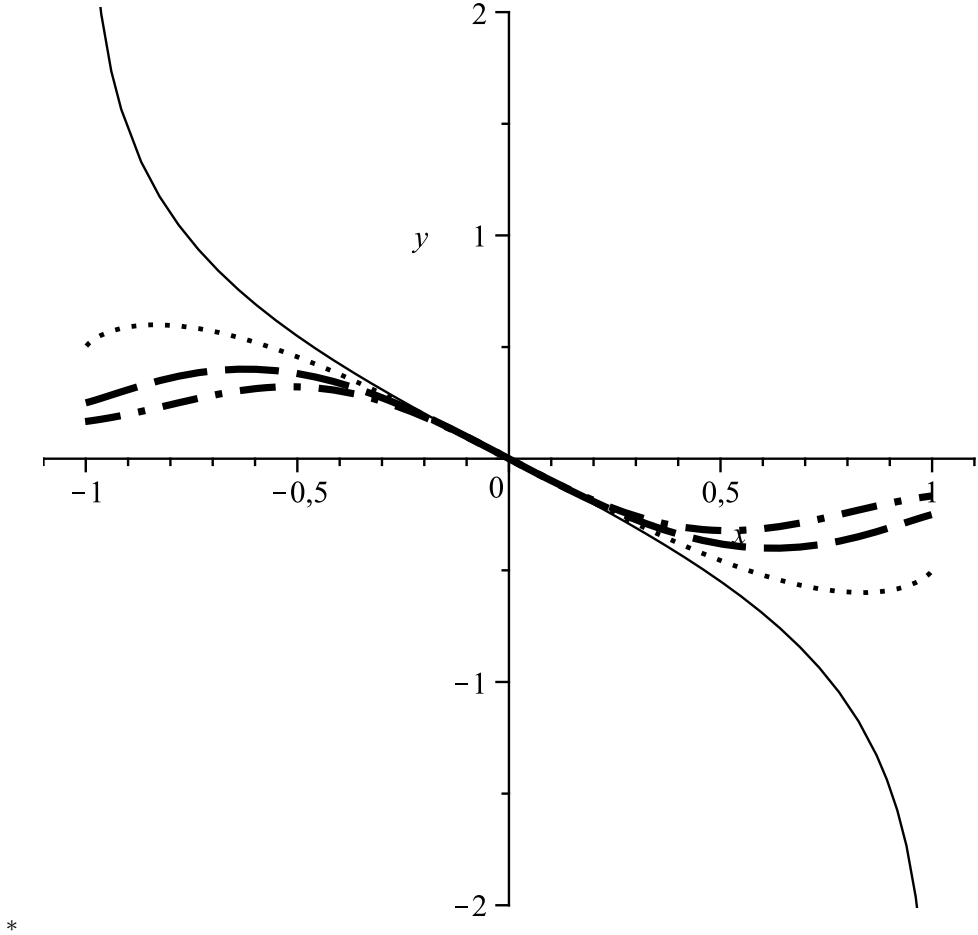}}}
\vspace*{0.4cm}
\caption{Solutions for (Eq. 13) for zero and positive integer  $a/2\epsilon $ values. Thin solid line is for 0, thin dotted line is for 1, thick 
dashed line is for 2 and thick dash dotted line is for 3. }
\label{elso}       
\end{figure}
 
\begin{figure}*  
\scalebox{0.5}{
\rotatebox{0}{\includegraphics{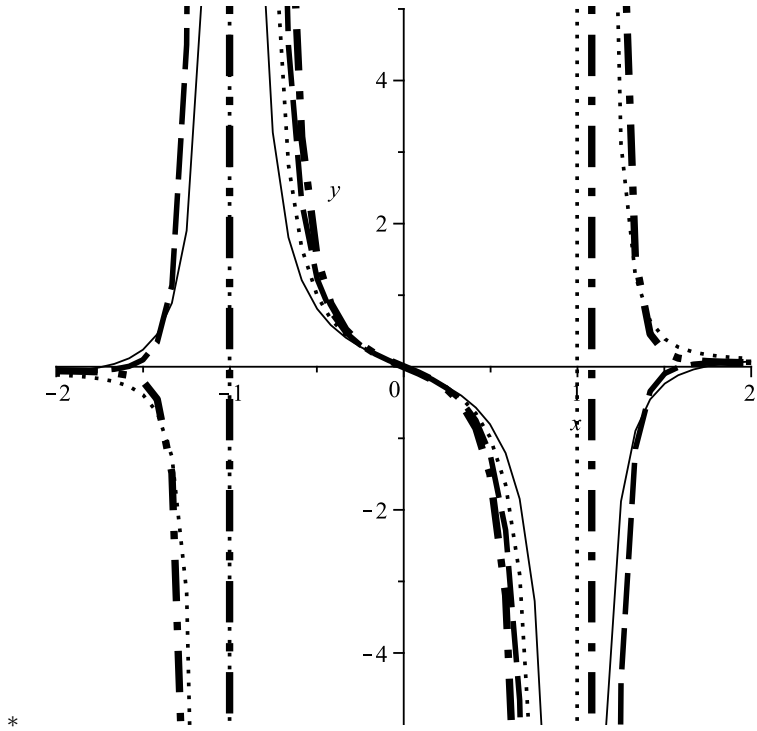}}}
\caption{Solutions for (Eq. 13) for negative integer  
$a/2\epsilon $ values. Thin solid line is for -1, thin dotted line is for -2, thick 
dashed line is for -3 and thick dash dotted line is for -4}	
\label{kettes}       
\end{figure}

\begin{figure}*
\scalebox{0.65}{
\rotatebox{0}{\includegraphics{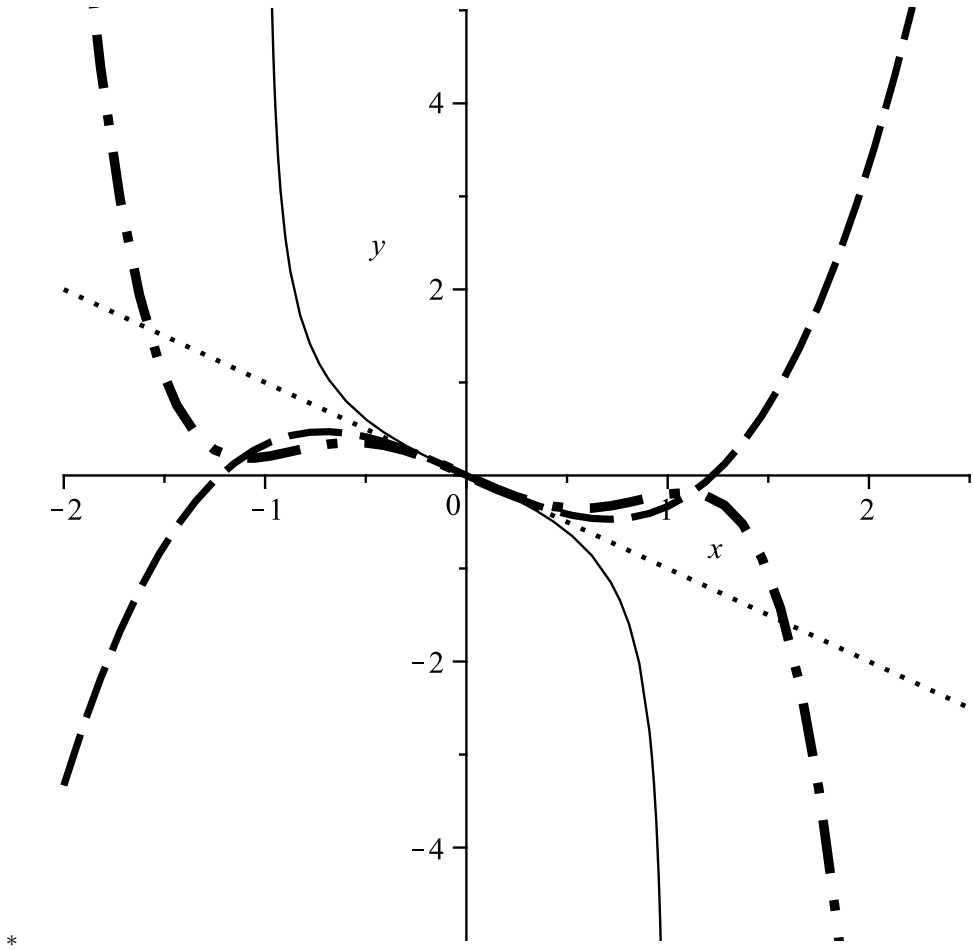}}}
\vspace*{0.4cm}
\caption{Solutions for (Eq. 13) positive half-integer $a/2\epsilon$ values.  Thin solid line is for 1/2, thin dotted line is for 3/2, thick dashed line is for 5/2 and thick dash dotted line is for 7/2.   }	 %
\label{elso}       
\end{figure}
 
\begin{figure}*  
\scalebox{0.5}{
\rotatebox{0}{\includegraphics{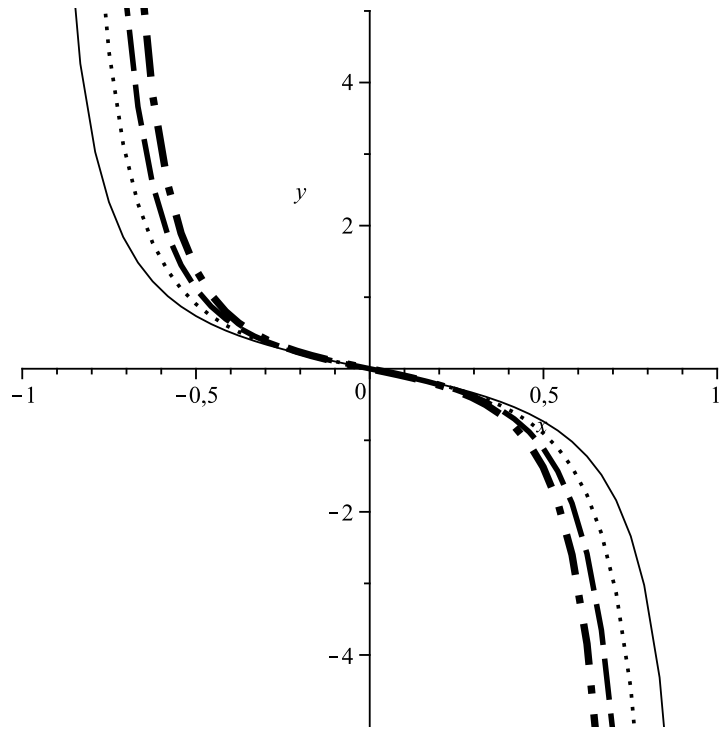}}}
\caption{Solutions for (Eq.13) the thin solid line is for 
$a/2\epsilon =-1/2$, thin dotted line is for -3/2, thick 
dashed line is for -5/2 and thick dash dotted line is for -7/2. }
\label{kettes}       
\end{figure}

 If the first integration constant  is not zero $(c_1 \ne 0 )$ than the following first ODE should be integrated
\eq
\epsilon(\eta^2 f(\eta))' - a\eta f(\eta) = f'(\eta) + c_1
\eqe
after some algebra the solution can be written in the following form
\eq  
f(\eta) = \left[ c_1 (signum (\epsilon \eta^2 -1 )^{ -\frac{a}{2\epsilon}})
 (-signum(\epsilon \eta^2 -1 ))^{\frac{a}{2\epsilon}}) \cdot \eta \cdot  _2 F_1\left( \frac{1}{2}, \frac{a}{2 \epsilon}; \frac{3}{2}; \epsilon \eta^2 \right) +c_2 \right]
\cdot (\epsilon \eta^2 -1 )^{ \frac{a}{2\epsilon}-1}
  \eqe
Note, that the former solution is still there independently and as a kind of form factor. 
Where $ _2 F_1 $ is the hypergeometric function \cite{abr,bate,whit,magnus}. It is well known that many elementary functions can be expressed via the hypergeometric function.  Fortunately, in our case if $\frac{a}{2\epsilon} $ is an integer or a half-integer the hypergoemtric series breaks down to a finite 
sum of terms. 
There are four basic cases which has crucial importance: 
\eq 
 \frac{a}{2\epsilon} = 0,  \hspace{0.5cm}  
_2 F_1\left(0,\frac{1}{2};\frac{3}{2};\epsilon \eta^2\right) = 1 
\eqe
\eq
\frac{a}{2\epsilon} = 1,  \hspace{0.5cm}  _2 F_1\left(1,\frac{1}{2};\frac{3}{2};\epsilon \eta^2\right) = 
\frac{1}{2{\sqrt{\epsilon}}\eta} \ln \left( \frac{1+ {\sqrt{\epsilon}}\eta} { 1- {\sqrt{\epsilon}}\eta}  \right).
\eqe 
for half-integer values 
\eq
\frac{a}{2\epsilon} = \frac{1}{2},  \hspace{0.5cm} _2 F_1\left(\frac{1}{2},\frac{1}{2};
\frac{3}{2};\epsilon \eta^2\right) = \frac{ arccos ({\sqrt{\epsilon\eta^2}}  )}
{{\sqrt{\epsilon\eta^2}}} 
\eqe
\eq
 \frac{a}{2\epsilon} = \frac{3}{2},  \hspace{0.5cm} 
_2 F_1\left(\frac{3}{2},\frac{1}{2};\frac{3}{2};\epsilon \eta^2\right) = \frac{1}{\sqrt{(1-\epsilon\eta^2)}}.
\eqe  
With the following recursion relation all the other cases can be evaluated
\eq
(c-a) _2 F_1(a-1,b;c;z) + (2a-c-az+z) _2 F_1(a,b;c;z) + a(z-1) _2 F_1(a+1,b;c;z) = 0.
\eqe
Even for negative parameters there are closed relations 
\eq
\frac{a}{2\epsilon} = -1, \hspace{0.3cm}  \epsilon > 0,  \hspace{0.5cm}  
_2 F_1\left(-1,\frac{1}{2};\frac{3}{2};\epsilon \eta^2\right) = 1 -\frac{\epsilon \eta^2}{3}
\eqe
\begin{eqnarray}
\frac{a}{2\epsilon} = -\frac{1}{2}, \hspace{0.3cm}  \epsilon > 0,  \hspace{0.5cm}  _2 F_1\left(-\frac{1}{2},\frac{1}{2};\frac{3}{2};\epsilon \eta^2\right) =  \nonumber \\ 	
\hspace{-3cm}   \frac{1}{2}
\left\{ (1/2 - \epsilon \eta^2)\frac{arcsin ({\sqrt{\epsilon}\eta})} {{\sqrt{\epsilon}\eta}}  
-\frac{\epsilon\eta^2-1 }{2(1- \epsilon\eta^2)^{1/2}  }   \right\}.   
\end{eqnarray}

Let's consider $c_1 =1$ and $c_2 =0$ solutions and analyze it in details. 
We will examine 4 different cases, (from physical reasons the propagation velocity $\epsilon$ always have to be positive, we take the plus unity value, and changing the 'a' 
parameter). 
For the first case let's take $ a/2\epsilon \ge 0 $ with integer values. The resulting curves are presented on Figure 1. We can see, that for positive integer $a/2\epsilon$ values the solution is compact but have a finite jump at  $\pm 1$.  
Figure 2 shows the same results but when $a/2\epsilon$ is equal to -1,-2,-3 and -4.
All figures are continuous at the [-1:1] and have singular asymptotic values at the boundaries.  
Figure 3 presents the results for positive half-integer $a/2\epsilon$  values. 
All the curves are odd functions. 
Figure 4 presents curves for negative half integer $a/2\epsilon$ values. 
All the results are odd functions and have asymptotes at $\pm 1$. 

Figure 5 presents the self-similar solution 
\eq  
T(x,t)=  \frac{1}{t}\left[ (signum (\epsilon (x/t)^2 -1 )^{ -\frac{a}{2\epsilon}})
 (-signum(\epsilon (x/t)^2 -1 ))^{\frac{a}{2\epsilon}}) \cdot \left(\frac{x}{t}\right) 
\cdot  _2 F_1\left( \frac{1}{2}, \frac{a}{2 \epsilon}; \frac{3}{2}; \epsilon (x/t)^2 \right)  \right]
\cdot \left[\epsilon \left(\frac{x}{t} \right)^2 -1 \right]^{ \frac{a}{2\epsilon}-1}
\label{presented}
  \eqe 
for $a/2\epsilon = 2$. 
We can see the discontinuous jump at the x=t line. For large x and t the solutions 
goes to zero.  \\

\begin{figure}*  
\scalebox{0.5}{
\rotatebox{0}{\includegraphics{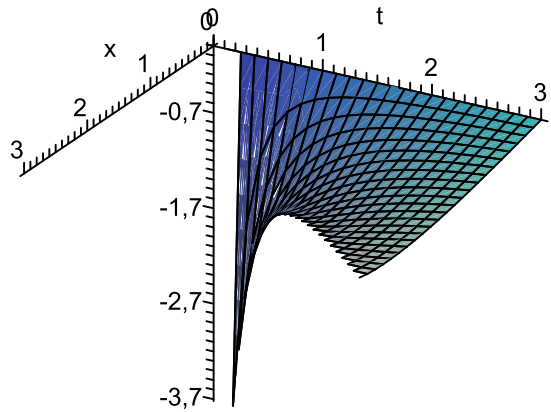}}}
\caption{The self-similar T(x,t) solution (Eq. \ref{presented})  for $a/2\epsilon = 2$.}
\label{kettes}       
\end{figure}

In the following we present solutions where alpha is not unity, we consider 
the  $\alpha = -2$ and $\beta = +1 $ values. (For arbitrary alphas the solutions are similar to this case but the details are much more complicated to understand.)
Now (Eq. 4) becomes  
 \eq
f''(\eta)[\epsilon\eta^2 - 1 ] -
f'(\eta) \eta[2\epsilon + a]  +
2f(\eta)[\epsilon + a] = 0.
\label{final}
\eqe

For general values of $\epsilon$ and $a$ the solutions are the
following:
\eq
f(\eta) =  c_1 P_{\frac{a}{2\epsilon}-1}^{\frac{a}{2 \epsilon}+2}
(\sqrt{\epsilon}\eta ) (\epsilon \eta^2-1)^{\frac{a}{4\epsilon}+1},    +
c_2 Q_{\frac{a}{2\epsilon}-1}^{\frac{a}{2\epsilon}+2}
(\sqrt{\epsilon} \eta) (\epsilon \eta^2-1)^{\frac{a}{4\epsilon}+1},
\label{legendre}
	\eqe
where $P_{a/2\epsilon-1}^{\frac{a}{2\epsilon}+2}(\sqrt{\epsilon} \eta) $  and
      $Q_{a/2\epsilon-1}^{\frac{a}{2\epsilon}+2}(\sqrt{\epsilon} \eta) $ 
are the associated Legendre functions
 of the first and second kind \cite{abr,bate,whit,magnus,hobs,mac}.
This result can be obtained by tedious calculations transforming
to the standard form of the Legendre differential equation Kamke \cite{kamke} or by
using Maple 9.1 program package.
The first and second arguments of P and Q  are the order $\nu = a/2\epsilon-1  $
and the  degree $\mu = a/2\epsilon +2 $ of the Legendre function, which may take unrestricted
real values. The variable $ y = \sqrt{\epsilon} \eta $ may have complex values es well.

It is well known that if the order and the degree are integer numbers ($\nu = l \epsilon N $,
$\mu = m \epsilon N$) and the order is larger that the degree ($\nu > \mu$) the
associated Legendre functions of the first kind became the associated Legendre polynomials.
These polynomials span a Hilbert space with the following orthonormalization condition:
\begin{eqnarray}
 \int_{-1}^{+1}P_{l_1}^m(x) P_{l_2}^m(x) dx = 0 \quad \quad  (l_1 \ne l_2 ) \nonumber \\
 \quad \quad \quad       = \frac{2}{2l_1+1}\frac{(l_1+m)!}{(l_1-m)!} \quad  \quad l_1 = l_2 )
\end{eqnarray}
Similar relations are available for the second order associated
Legendre polynomials $Q_l^m(x)$ as well.

If the order and the degree are non-integer values than the Legendre functions can 
be evaluated with the hyperbolic functions\cite{abr,bate,whit,magnus}, in our case the formulas are the following:

\eq
P_{\frac{a}{2\epsilon}-1}^{\frac{2}{2\epsilon}+2}(\sqrt{\epsilon} \eta) =
\frac{(\sqrt{\epsilon}+1)^{\frac{a}{4\epsilon}+1}} 
{(\sqrt{\epsilon}\eta-1)^{\frac{a}{4\epsilon}+1}  \Gamma(-\frac{a}{2\epsilon}-1) } 
\times _2 F_1\left(\frac{a}{2\epsilon},-\frac{a}{2\epsilon}+1; -\frac{a}{2\epsilon}-1; 
 \frac{1}{2} -\frac{ {\sqrt{\epsilon}} \eta}{2} \right) \nonumber  \\ 
\label{conf}
\eqe
and

\begin{eqnarray}
Q_{\frac{a}{2\epsilon}-1}^{\frac{2}{2\epsilon}+2}(\sqrt{\epsilon} \eta) =
\frac{1}{2}e^{\frac{I(4\epsilon+a)\Pi}{2\epsilon}}(\epsilon+a)
2\Gamma \left(\frac{a}{2\epsilon}-1 \right)  
\times _2 F_1\left(\frac{a}{2\epsilon}+\frac{3}{2},\frac{a}{2\epsilon}+1; \frac{a}{2\epsilon}+\frac{1}{2}; 
 \frac{1}{2} -\frac{ {\sqrt{\epsilon}}\eta}{2} \right) \times  \nonumber \\
2^{(\frac{a}{2\epsilon} +1)} ({\sqrt{\epsilon}}\eta-1)^{\frac{a}{4\epsilon}-1} 
({\sqrt{\epsilon}} \eta+1)^{\frac{a}{4\epsilon}-1}/\left[\epsilon (\sqrt{\epsilon}\eta)^{\frac{a}{\epsilon}+2}\right]. 
 \end{eqnarray}
The function $ _2 F_1(...) $ is the
hypergeometric function again, and $\Gamma(...)$
is the gamma function \cite{abr,bate}.
In our case, both the hypergeometric functions has a large degree of symmetry 
hence, the series expansion breaks down after some terms
\eq
_2 F_1\left(\frac{a}{2\epsilon},-\frac{a}{2\epsilon}+1; -\frac{a}{2\epsilon}-1; 
 \frac{1}{2} -\frac{ {\sqrt{\epsilon}} \eta}{2} \right)=
4\frac{(1+ \epsilon\eta^2 + \eta^2a)\epsilon)(\frac{1}{2}+\frac{{\sqrt{\epsilon}}\eta}{2})}
{(1+2\sqrt{\epsilon}\eta + \epsilon\eta^2)(2\epsilon +a)}
\eqe
and 
\eq
_2 F_1\left(\frac{a}{2\epsilon}+\frac{3}{2},\frac{a}{2\epsilon}+1; \frac{a}{2\epsilon}+\frac{1}{2}; 
 \frac{1}{2} -\frac{ {\sqrt{\epsilon}} \eta}{2} \right) = 
\frac{\epsilon(1+\epsilon\eta^2 + a\eta^2 )}{(\epsilon+a)(\epsilon\eta^2-1)
\left(\frac{\epsilon\eta^2-1}{\epsilon\eta^2} \right)^{\frac{a}{2\epsilon}+1} }.
\eqe
Now, after some algebra the total regular solution can be obtained as follows
($c_1= 1, c_2 =0$)

\eq
f(\eta) = P_{\frac{a}{2\epsilon}-1}^{\frac{a}{2\epsilon} +2}(\sqrt{\epsilon}\eta)(\epsilon
\eta^2 -1)^{\frac{a}{4\epsilon} +1}  = \frac{2\epsilon
(\sqrt{\epsilon}\eta+1)^{\frac{a}{2\epsilon}+1}(1+\eta^2(\epsilon+a)) } 
{\Gamma \left( \frac{-a}{2\epsilon} -1\right) (2\epsilon +a) }
\label{abraz}
\eqe

\begin{figure}*  
\scalebox{0.5}{
\rotatebox{0}{\includegraphics{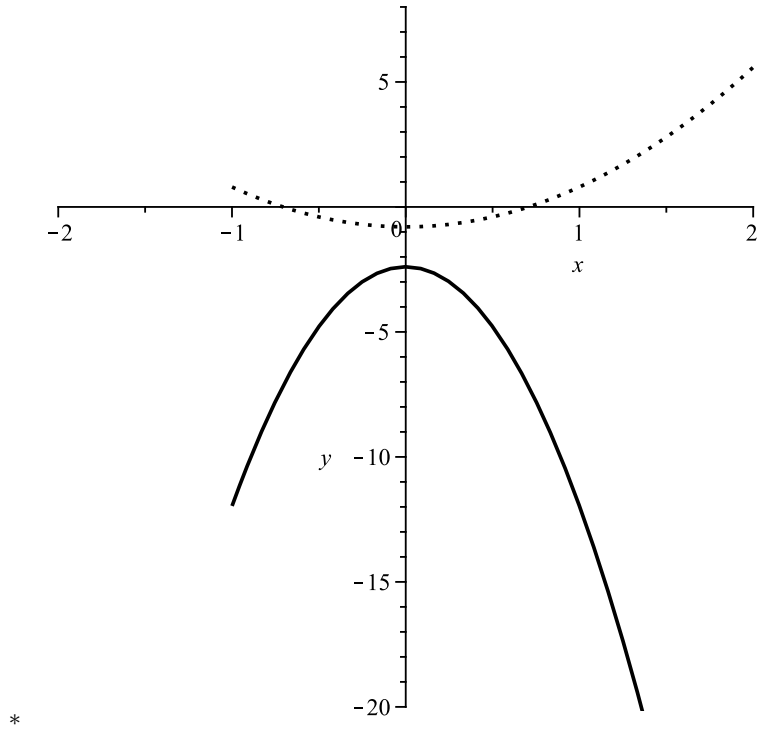}}}
\caption{Solutions of (Eq. \ref{abraz}), the solid line if for $a/2\epsilon = 3/2$ and 
the dotted line is for $a/2\epsilon = -3/2$}.
\label{kettes}       
\end{figure}

Analyze now the results, which means four different cases again. 
($a/2\epsilon$ is positive/negative integer, or $a/2\epsilon$ positive/negative half-integer).

 If 'a' is positive or negative odd number, then $a/2\epsilon$ is a half integer number 
and the domain of $(\sqrt{\epsilon}\eta+1)^{a/2\epsilon+2}$ is for $\eta > -1$ 
which means that the $ (1+\eta^2(\epsilon+a))$ parable is only considered for $\eta > -1$. 
Such solutions can be seen on Fig 6. 
 
If 'a' is an even number then the cases are a bit more difficult. 
From the properties of the Gamma function is clear that if $a/2\epsilon$ is a negative integer, (a is a positive even number) than $f(\eta)$ is zero. 
So 'a' have to be negative and even.  
From the weight factor of (Eq. 23)   $(\epsilon\eta^2-1)^{\frac{a}{4\epsilon}+1}$  we can see that if a is divisible with 4 than the power of $(\epsilon\eta^2-1)$ is an integer, 
and the range is the whole real axis, however if a is negative even but not divisible with 
4 than the power of $(\epsilon\eta^2-1)$  is a half-integer, which means the range is not 
continuous on the whole real axis. 
This property together with the Legendre function gives a cut at $\eta =-1$. 
Such solutions are presented on Fig 7.

\begin{figure}*  
\scalebox{0.5}{
\rotatebox{0}{\includegraphics{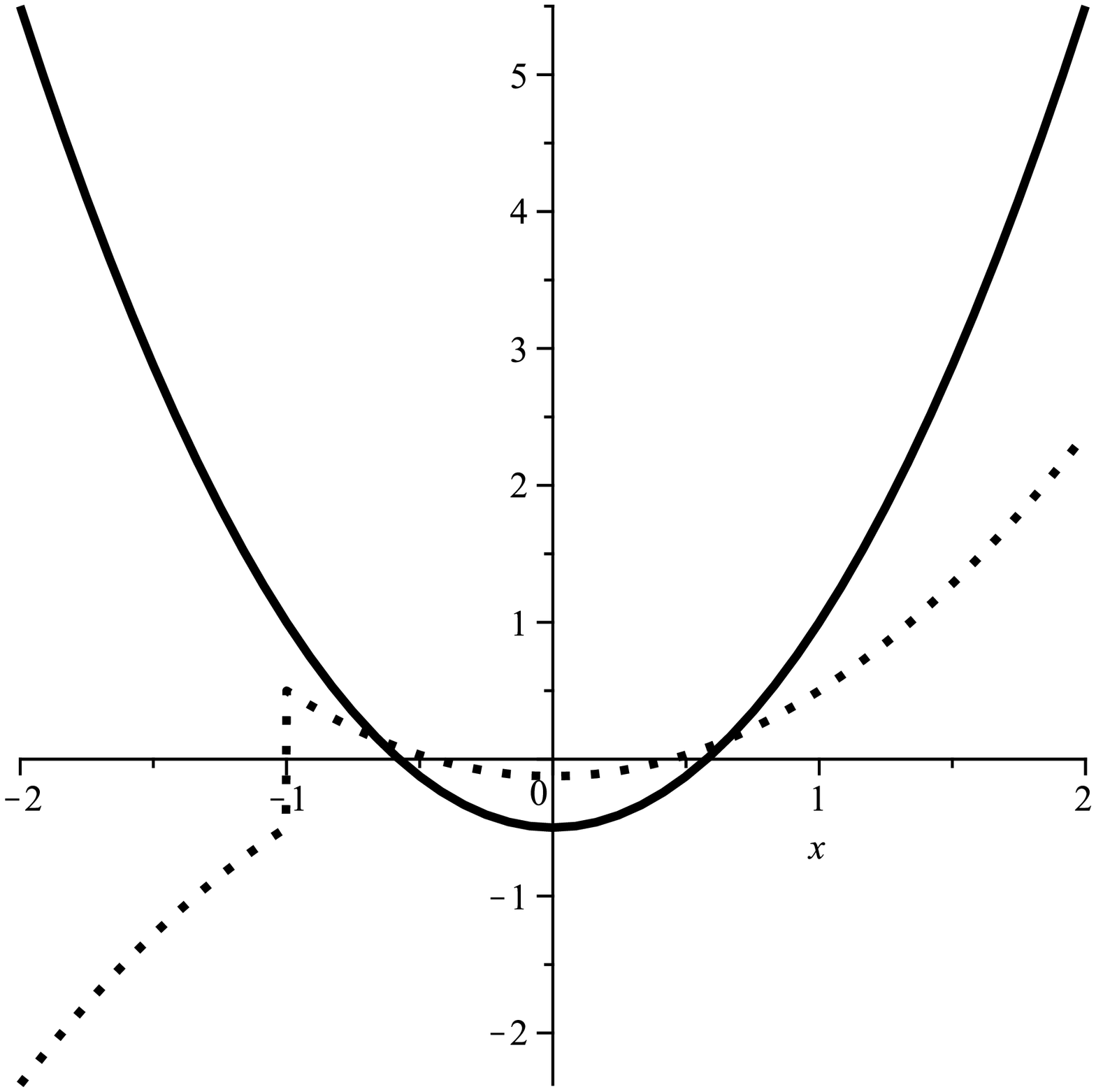}}}
\caption{Solutions of (Eq. \ref{abraz}), the solid line if for $a/4\epsilon+1 = -1$ and 
the dotted line is for $a/4\epsilon +1= -\frac{1}{2}$.}
\label{kettes}       
\end{figure}

The total self-similar solution (Eq. 3) is presented for $a=-2, \epsilon=1$
\eq
T(x,t) = t^2f(x/t) = t^2 P_1^1\left( \frac{x}{t} \right) \sqrt{\left[ \left(
\frac{x}{t}\right)^2-1 \right] }  
\eqe
of Figure 8. The solution is a nice  continuous function.  

\begin{figure}*  
\scalebox{0.5}{
\rotatebox{0}{\includegraphics{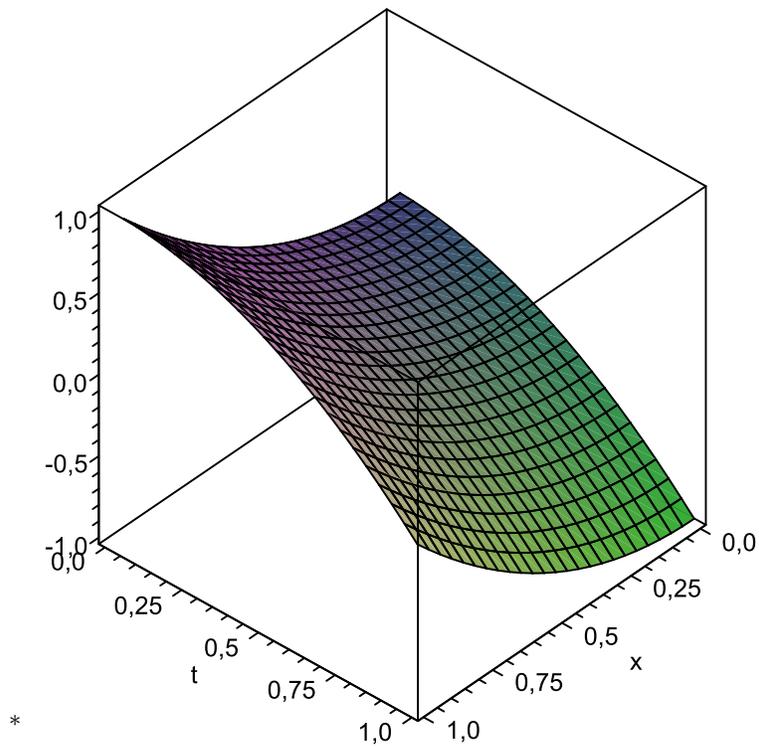}}}
\caption{The self similar-solution for $a= -2, \epsilon =1$} 
\label{P3dim}       
\end{figure}

For the irregular solution  ($c_1=0, c_2 =1$)  the solution formula is however a bit more complex, however still four different cases have to be analyzed [$(a/2\epsilon)$ is positive/negative integer or positive/negative half-integer value]. 
The irregular Q Legendre function is not defined for any negative (integer and half-integer values also) order values ($a/2\epsilon-1$).      
The complex exponential factor in (Eq. 25) $ e^{\frac{I(4\epsilon+a)\pi}{2\epsilon}} $can induce pure real, pure imaginary or mixed solutions. 
E.g. for ${\frac{a}{4\epsilon}+1} =3/2$ or $ 7/2 $ the next formula is valid
\eq
i(\epsilon\eta^2-1)^{\frac{a}{4\epsilon}+1} = (1-\epsilon\eta^2)^{\frac{a}{4\epsilon}+1}
\eqe
which interchanges upper or lower boarders.

Figure 9 presents the results for $\epsilon =1, a=1,2,3,4$. 
The domain of the $ (\eta^2-1)^{5/4}Q_{\frac{-1}{2}}^{\frac{5}{2}}(\eta)$ is for 
$\eta < -1$ the range is also bounded from below. 
The second solution $ (x^2-1)^{3/2}Q_0^3(x) $ has a continuous range and domain with a 
jump at $\eta = -1$.   
Solution $(\eta^2-1)^{7/4}Q_{\frac{1}{2}}^{\frac{7}{2}}(\eta)$ is very similar to the first 
solution. The domain of the last curve $(\eta^2-1)^2Q_1^4(\eta) $ is the $-1<\eta<1$ opened interval and the function has singular values at the boarders. 

\begin{figure}*  
\scalebox{0.5}{
\rotatebox{0}{\includegraphics{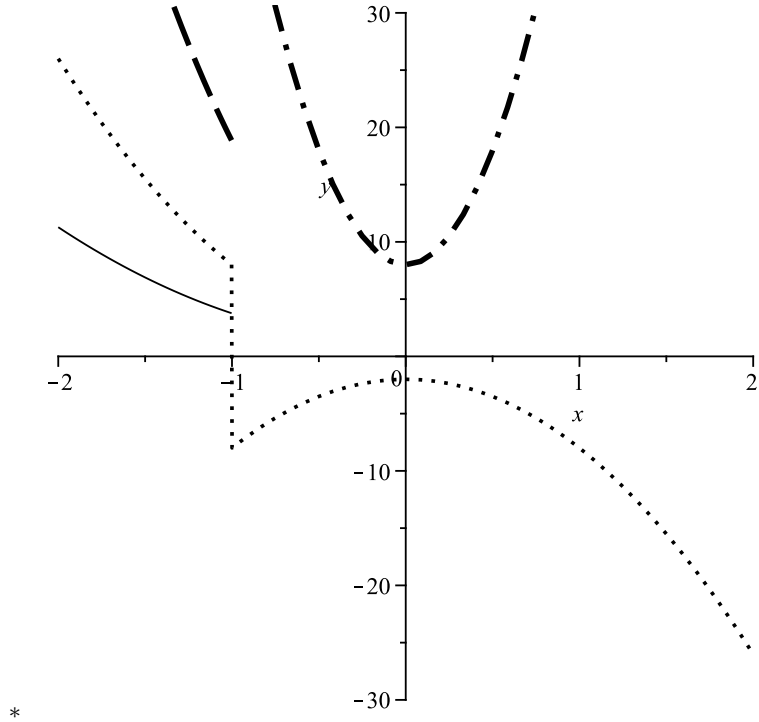}}}
\caption{ The results for the irregular Legendre function the solid line if for $a=1$ and 
the dotted line is for $a=2$ the dashed line is for $a=3$ and the dot-dashed is for $a=4$.}
\label{kettes}       
\end{figure}

At this point, we mention than all kind of solutions can be written in the form 
of the product of two travelling waves propagating in opposite directions. If we 
insert  $ c^2 = 1/\epsilon $ (the wave-propagation speed) into an irregular solution 
e.g. 
\eq
T(x,t) =  t^2  Q_1^4 \left(\frac{ \sqrt{\epsilon}x}{t} \right) \left[ \epsilon\left(  \frac{x}{t} 
\right)^2 -1 \right]^{2}
\label{Q3d}
\eqe
after some algebraic manipulation we get
\eq
T(x,t) =  t^2 Q_1^4 \left( \frac{x}{ct} \right) \> U(x-ct) U(x+ct)
\eqe
where $U(x \pm ct)= (x \pm ct)^{2}$.
Which means a distorted wave solution with a non-trivial weight function.
Figure 10 presents a self-similar irregular solution (Eq. \ref{Q3d}). 
Note, the cut at the $x=t$ line. 
We think that after this kind of analysis all the properties of the solutions 
are examined and understood.   

\begin{figure}*  
\scalebox{0.5}{
\rotatebox{0}{\includegraphics{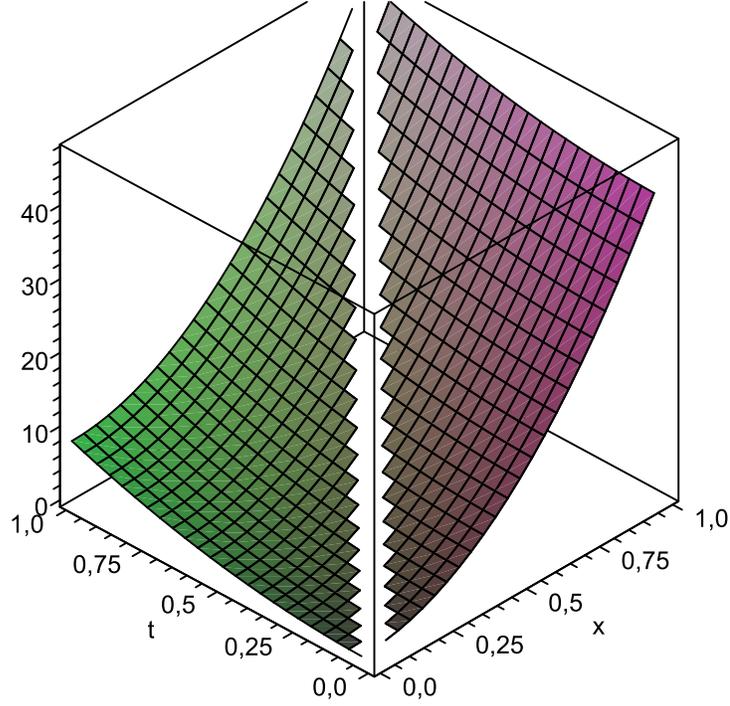}}}
\caption{ The graph of the irregular self-similar solution of Eq. \ref{Q3d}.}
\label{kettes}       
\end{figure}
 
\vspace*{1cm}
\hspace*{6cm} ---------------  \\

Let's consider now the 2 dimensional telegraph-type equation of the following form: 
 \begin{equation}\label{5}
\epsilon\frac{\partial^2 S(x,y,t)}{\partial t^2} + \frac{a}{t} \frac{\partial S(x,y,t)}{ \partial t}
= \frac{\partial^2 S(x,y,t)}{\partial x^2} + \frac{\partial^2 S(x,y,t)}{\partial y^2},  
 \end{equation}
 
To avoid further confusion we are looking for the solution in the form of 
\eq
S(x,y,t)=t^{-\alpha}g\left(\frac{x+y}{t^\beta}\right):=t^{-\alpha}g(\eta). 
\eqe
We may consider this as a the $L^1$ vector norm as well. 

The idea behind this Ansatz is the exchange symmetry over the spatial coordinates: 
\eq
\frac{\partial^2 S(x,y,t)}{\partial x^2} = \frac{\partial^2 S(x,y,t)}{\partial y^2} = t^{-\alpha}
g''(\eta)t^{-2\beta}. 
\eqe
 where prime denotes differentiation with respect to $\eta$.

If we use the condition that $\beta = +1 $  we get the most general ordinary differential 
equation: 
\begin{equation}\label{13}
    g''(\eta) [\epsilon\eta^2-2 ] - 
g'(\eta) \eta(a -3\epsilon) +
g(\eta) \alpha( \alpha\epsilon  + \epsilon -a) = 0.
\end{equation}
The solution is the following: 
\begin{eqnarray}
g(\eta) =  c_1(\epsilon \eta^2 -2)^{\frac{a}{4\epsilon}-\frac{1}{4}}
P_{\omega}^{\frac{a}{2\epsilon}-\frac{1}{2}}\left[{\sqrt{\frac{\epsilon}
{2}}}\eta\right]  + \nonumber \\ 
c_2(\epsilon \eta^2 -2)^{\frac{a}{4\epsilon}-\frac{1}{4}}
Q_{\omega}^{\frac{a}{2\epsilon}-\frac{1}{2}}\left[{\sqrt{\frac{\epsilon}{2}}}\eta\right] 
\end{eqnarray}
where P and Q are the associated Legendre functions, and  \\ 
\eq
\omega = \frac{\sqrt{(-4\alpha + 4 - 4\alpha^2)\epsilon^2 +4a(\alpha -1)
\epsilon +a^2} -\epsilon}
{2\epsilon}
\eqe
If $\alpha = +1 $ than $|a| > |2\epsilon|$ is a constrain for a real $\omega$ number which is the 
order of the Legendre function.  
If we consider the $-4\alpha + 4 - 4\alpha^2 = 0$ condition than it comes out 
$\alpha_1 = -\sqrt{5}/2 - 1/2 $ or 
$\alpha_2 = \sqrt{5}/2 - 1/2 $. 
This means for $\alpha_1$ the $a<0$ and for $\alpha_2$ the $a>0$ 
conditions have to be fulfilled to have a real order for the Legendre function. 
It is interesting to mention that the number $ 
{\frac{1 +{\sqrt5}}{2}} \approx  1.61803. $ is the golden ratio.

If $\beta = +2 $ the ODE and the solutions are the followings: 
\begin{equation}\label{13}
    g''(\eta) [\epsilon\eta^2-2 ] - 
g'(\eta) \eta(a -3\epsilon) +
2g(\eta)( \epsilon +a) = 0.
\end{equation}
\eq
g(\eta) =  c_1(\epsilon \eta^2 -2)^{\frac{a}{4\epsilon}-1/4}P_
o^{a/2/\epsilon-1/2}(\sqrt{\epsilon/2}\eta) + 
c_1(\epsilon \eta^2 -2) ^{\frac{a}{4\epsilon}-1/4}Q_o^{a/2/\epsilon-1/2}
(\sqrt{\epsilon/2}\eta) 
\eqe
where o has a bit simpler form: 
\eq
o = \frac{\sqrt{-4\epsilon^2 - 12a\epsilon +a^2} -\epsilon}{2\epsilon}
\eqe
 to have real number for the order of the associated Legendre function 
the "a" and $\epsilon$ values are not independent from each other, and the $a/2\epsilon$ 
relation became a bit more complicated. However, the structure of the solutions 
remains the same, like for the 1 dimensional case. Some solutions are continuous on the 
whole real axis and some of them have a bonded range from above or from below. \\ 

As for the more usual Euclidean norm we may consider the following Ansatz as well
\eq
S(x,y,t)=t^{-\alpha}g\left(\frac{\sqrt{ x^2+y^2}}{t^\beta}\right):=t^{-\alpha}g(\eta)
\eqe
where $\sqrt{x^2+y^2}$ is the usual distance between the two points, which is never 
negative. 
After having done the usual derivation we got $\beta =1$  and the 
following ODE 
\eq
g''(\eta)(\epsilon\eta^2 -1) + g'(\eta)([2\epsilon\alpha + 2\epsilon -a]\eta - 1/\eta) 
+ g(\eta)(\epsilon\alpha^2+\epsilon\alpha -\alpha a) = 0  
\eqe
the most general solution is the following: \\
\begin{eqnarray}
g(\eta) = c_1\cdot _2F_1\left(\frac{\alpha}{2},\frac{(1+\alpha)\epsilon-a}{2\epsilon}; \frac{2\alpha\epsilon+\epsilon-a}{2\epsilon}; 1-\epsilon\eta^2 \right)  +  \nonumber \\
c_2 \cdot _2 F_1\left(1-\frac{\alpha}{2},\frac{e+a-\alpha\epsilon}{2\epsilon}; \frac{(3-2\alpha)\epsilon+a}{2\epsilon}; 1-\epsilon\eta^2 \right) (\epsilon\eta^2-1)^{\frac{a+(1-2\alpha)\epsilon}{2\epsilon}}. 
\end{eqnarray}
Which means that even this Ansatz is not contradictory, and gives a reasonable 
ODE. Note, that the only basic difference between this ODE and the original one (Eq. 4) 
is the extra $1/\eta$ term after the first derivative of $g(\eta)$. 
From the $(\epsilon\eta^2-a)$ form factor of the second derivative we can see,
\cite{whit} that this ODE also has singular points at $\eta = \pm1$ for $\epsilon =1$ which is not special for wave-equations. If these singularities are relevant or can be eliminated   
depends on the relations of the parameters, (see Fig. 9. and explanations there).    
Here, we can find large number of analytic solutions again (for integer/half integer 
$\alpha, \epsilon$ and a), which are similar to (Eq. 14 - 20). 

We hope that with the help of these two mentioned models we can analyze 2 dimensional dissipative flows int the future where the the original telegraph equation plays an 
important role \cite{wilh}.    

\vspace*{1cm}
\hspace*{6cm} ------------------------ \\ 

At last we investigate the one-dimensional telegraph-type equation with a 
general source term 
 
\begin{equation} 
\epsilon\frac{\partial^2U(x,t)}{\partial t^2} + \frac{a}{t} \frac{\partial U(x,t)}{ \partial t} - \frac{\partial^2 U(x,t)}{\partial x^2} + s(x,t)U(x,t) = 0	 \label{source}
 \end{equation}
 
We are looking for solution of (\ref{source}) of the form
\eq
U(x,t)=t^{-\alpha}h\left(\frac{x}{t^\beta}\right):=t^{-\alpha}h(\eta).
\eqe
(To avoid confusion we use different letters for the basic variable, and for the Ansatz as well.)
After the standard derivation and parameter investigation it comes out that $\beta = 1$ is a necessary condition. The most general ODE without a source term is still (Eq. 4.). 
Note, that any kind of source term (without derivation) gives additional 
terms for the last parenthesis which should be a function of $\eta$.  
There are two important source terms in physics, the first is the harmonic 
oscillator. 
So $s(x,t) = Dx^2$ where D is the stiffness of the oscillator. This automatically fix $\alpha $ to +2. 
Our ODE is now: \\ 
\begin{equation}\label{13}
    h''(\eta) [\epsilon\eta^2-1 ]  + 
h'(\eta) \eta [6\epsilon-a] -
h(\eta) [-6\epsilon +2a +D\eta^2]  = 0.
\end{equation}
the solution is the 
\begin{eqnarray}
 h(\eta) = c_1\cdot HeunC \left( 0,-\frac{1}{2}, \frac{-4\epsilon+a}{2\epsilon} , 
- \frac{D}{4\epsilon^2}, \frac{10\epsilon -3a}{8\epsilon},
\epsilon\eta^2 \right)\cdot 
\left[\epsilon \eta^2 -1\right]^{\frac{a-4\epsilon}{2\epsilon} } 
+ \nonumber \\
c_2\cdot HeunC \left(0,\frac{1}{2} , 
\frac{-4\epsilon+a}{2\epsilon} , -
\frac{D}{4\epsilon^2}, \frac{10\epsilon -3a}{8\epsilon},
\epsilon\eta^2 \right) \eta   
\left[\epsilon \eta^2 -1\right]^
{\frac{a-4\epsilon}{2\epsilon}}
\end{eqnarray} 
  HeunC function which has three singularities. The regular ones are in $\mp \frac{1}{\sqrt{\epsilon}} $ and the irregular is in infinity. 
There are different kind of Heun functions, and all kind can be evaluated 
with a generalized power series, or for special arguments represents different kind of special functions like the 
hypergeometric functions, Legendre functions, Bessel functions, 
Gegenbauer polinoms, Kummer functions and so on. 
Unfortunately, for the above given parameters, the Heun function does not represent any kind of special functions, but the series expansion is still valid. 
\begin{eqnarray}
HeunC \left( 0,-\frac{1}{2}, \frac{-4\epsilon+a}{2\epsilon} , 
- \frac{D}{4\epsilon^2}, \frac{10\epsilon -3a}{8\epsilon},
\epsilon\eta^2 \right) = 1 + \left(\epsilon -\frac{a}{2}\right)\eta^2 \nonumber \\ 
- \left(\frac{a\epsilon}{12} -  \frac{a^2}{24} - \frac{c}{12}\right)\eta^4 \left(-\frac{a^3}{240} +\frac{(4\epsilon^2+2c)a}{240} 
-\frac{\epsilon c}{20} \right)\eta^6  +O(\eta^8)  
\end{eqnarray}

The other important solutions is the $s(x,t) = q/x$ case, which is the Coulomb potential 
 where q is the electrical charge . This automatically fix $\alpha = -1 $. 
Now the ODE is the following 
\begin{equation}\label{14}
    h''(\eta) [\epsilon\eta^2-1 ]  - 
h'(\eta) \eta a -
h(\eta) [-a + q/\eta]  = 0.
\end{equation}
The solution is 

\begin{eqnarray}
 h(\eta) = c_1\cdot HeunG \left( 2, \frac{q\epsilon +a\sqrt{\epsilon}}{\epsilon^{3/2}},-1 , 
-\frac{a}{\epsilon}, -\frac{a}{2\epsilon},0, 
\sqrt{\epsilon}\eta +1 \right) + \nonumber \\
c_2 \left[ \epsilon\eta^2-1 \right]^{\frac{a}{4\epsilon}}
\left[-\sqrt{\epsilon}\eta-1 \right]^{\frac{4\epsilon +a}{4\epsilon}}
\left[-\sqrt{\epsilon}\eta+1 \right]^{-\frac{a}{4\epsilon}} \cdot \nonumber \\
HeunG \left(2,\frac{q\epsilon^2-\frac{a\epsilon^{3/2}}{2}-\frac{\sqrt{\epsilon}a^2}{4}}
{\epsilon^{5/2}}, \frac{a}{2\epsilon}, 1-\frac{a}{2\epsilon}, 2 +\frac{a}{2\epsilon},0, 
\sqrt{\epsilon}\eta +1 \right)
\end{eqnarray} 
This solution has four singular points at 0 at  $\infty$ and at  $\mp 1/\sqrt{\epsilon}$.   
Without any physical restriction between the three parameters, further pure mathematical investigation has not much sense.    

We think that this model can help us to investigate further the features of the 
quantum telegraph equation \cite{sanch}, there the first time derivative has an additional 
complex unit parameter.    \\ 

{\it{In summary:}} \\ 
In the presented study we gave an in-depth analysis of a time-dependent telegraph-type 
equation for heat propagation. All the cases for different parameter ranges are carefully 
examined and analyzed. Large number of special functions came into play, which may serve as a strong hint for further relevance of this equation.    
We hope that our equation can help us to investigate two dimensional turbulent flows, 
or even some quantum mechanical problems.    
                                                                  
\end{document}